Antibody Consumption-Driven Dynamic Competition: A Systems Hypothesis for the Transition from Acute Immune Response to Post-Infection Sequelae

Shi Qiru*

LanzhouCity University

*Correspondence: shiqiru@lzcu.edu.cn

11 JiefangRoad, Anning District, Lanzhou, Gansu, China

Abstract: The mechanisms underlying the formation of post-infection sequelae are complex and remain controversial. This hypothesis integrates Bystryn's antibody feedback phenomenon and Imbiakha's immune cost theory, proposing for the first time a "Consumption-Driven Dynamic Competition of Antibody Clones" mechanism. This mechanism posits that the immune system may regulate the proliferation and differentiation of corresponding B cell clones by sensing and responding to the consumption rate of specific antibodies. This competition, driven by differences in consumption rates, might not only influence pathogen clearance efficiency and associated acute pathology during the antigen growth phase but also critically mediate the onset, development, and even resolution of post-infection sequelae during the antigen decay and homeostasis re-establishment phases. The proposed three-phase "consumption-driven dynamic competition" model provides a unified and dynamic explanatory framework for understanding the significant individual variability and dynamic evolution of post-infection immune outcomes (including the emergence and self-limitation of acute symptoms, the formation and persistence of chronic sequelae, and symptom fluctuations or resolution). It emphasizes not just specific molecules but the macroscopic dynamics of competition and selection within the immune system, offering a theoretical basis for exploring new intervention strategies for sequelae (e.g., by regulating the balance of antibody competition).

Keywords: Antibody, Consumption-driven, Dynamic competition, Immunodynamics, Post-infection sequelae

1. Introduction

Post-infection sequelae pose a persistent challenge to public health, with their complex mechanisms remaining controversial. Bystryn's antiserum feedback experiments suggested the possibility of compensatory regulation based on antibody levels [1], while Imbiakha proposed the concept that immune responses are accompanied by an "immunopathological cost" [2]. Although these theories offered new perspectives, indicating the existence of feedback or cost-tradeoff mechanisms within the immune system, their specific operational modes, especially at the level of dynamic regulation, lack in-depth investigation.

Recently, the complex sequelae following the COVID-19 pandemic [3], including abnormal antibody profiles (e.g., increased autoantibodies [4]), significant individual variability, and unpredictable disease course dynamics, have further highlighted the deficiencies in current understanding. This strongly suggests potential key gaps in existing theories: firstly, the crucial role of antibody "consumption rate" (reflecting dynamic clearance efficiency in vivo), rather than merely concentration, in driving B cell clonal selection might be overlooked; secondly, when exploring the causes of sequelae (especially those related to autoimmunity), research often focuses on mechanisms directly inducing self-reactivity, such as molecular mimicry or epitope spreading, potentially underestimating the importance of the "dynamic competition" effect arising from consumption rate differences among different B cell clones. Particularly during changes in antigen load, this competitive imbalance might provide a neglected window of opportunity for the development of sequelae.

To bridge these gaps, this study proposes the "Consumption-Driven Dynamic Competition of Antibody Clones" hypothesis. The core innovation lies in considering the "antibody consumption rate" as the key indicator driving B cell competition. This hypothesis aims to construct a unified dynamic framework to explain the efficiency of acute immune responses, related pathologies, and the onset, development, and resolution of chronic sequelae, offering a new systemic perspective for understanding and intervening in post-infection immune dysregulation.

2. Theoretical Basis

This hypothesis is based on the reinterpretation and extension of classical theories to construct the "consumption-driven dynamic competition" model.

2.1 Reinterpretation of Bystryn's Feedback Phenomenon: Consumption as a Driving Signal

The core inspiration for this hypothesis stems from a deep reinterpretation of Bystryn's classic antibody feedback experiments [1]. In these experiments, after selectively removing one type of antibody via exchange transfusion to cause a sharp drop in its serum level, a key phenomenon occurred: the level of the removed antibody began to rise sharply within just 24-48 hours. The core insight from this phenomenon is that its trigger is not a re-stimulation by antigen, but rather the antibody 'deficit' itself. This demonstrates the existence of an endogenous feedback loop that regulates antibody generation not by directly depending on antigen levels, but by sensing and responding to the concentration of circulating antibodies.We infer that the immune system is not merely sensing static antibody concentrations, but is more likely responding to the acute "deficit" or "consumption rate" of the antibody as a powerful signal to initiate compensatory generation. Based on this, we propose our core mechanism: the consumption rate of a specific antibody—reflecting its efficiency in binding to its target and being cleared—is a key signal driving the selection and amplification of the corresponding B cell clone.

This defines a "consumption-driven compensatory amplification and selection pathway." Crucially, the core signal triggering amplification via this pathway is the rapid consumption of the antibody itself, independent of classical antigen presentation and T cell help for activating the corresponding B cell clone. During rapid pathogen proliferation, antibodies that effectively bind the pathogen are quickly consumed. By sensing this high consumption rate, the immune system preferentially activates and expands the B cell clones producing these high-consumption-rate antibodies. This can rapidly increase the production of effective antibodies against the currently dominant pathogen epitopes in a short period, significantly increasing the proportion of these functional antibodies in the total antibody pool, thereby indirectly enhancing the "apparent" affinity or clearance efficiency of the antibody response towards the pathogen.

Furthermore, this pathway can efficiently synergize with the classical affinity maturation pathway: B cell clones preferentially expanded via consumption-driven signals (i.e., clones already preliminarily validated for target binding) can be more effectively recruited into germinal centers to participate in subsequent somatic hypermutation and affinity selection. This "pre-screening" mechanism greatly enriches the pool of effective clones entering the classical maturation pathway, providing "preferred seeds" for affinity maturation, thus significantly enhancing the overall efficiency and speed of high-affinity antibody production and accelerating the optimization process of the immune response.

2.2 Extension of Imbiakha's Immune Cost Theory: Competitive Imbalance and Sequelae Risk

Extending Imbiakha's view that the adaptive immune response itself can cause pathological damage (i.e., "immunopathological cost") [2], this hypothesis suggests that the "consumption-driven" mechanism inherently contains this risk. Although this mechanism promotes pathogen clearance by amplifying efficient antibodies during the antigen growth phase, its competitive advantage weakens as the pathogen-specific antibody consumption rate decreases during the antigen decay phase. At this point, if self-reactive B cell clones exist that are continuously consumed by binding to self-antigens (even if their absolute consumption rate is not high), their consumption-driven amplification signal might become relatively prominent in the context of reduced overall competitive pressure. If this potential amplification is not effectively constrained by immune tolerance mechanisms, it could lead to the inappropriate competitive success and expansion of self-reactive clones, ultimately initiating or sustaining post-infection sequelae. This provides an explanation based on dynamic competitive imbalance for why the process of clearing pathogens might induce the risk of sequelae.

3. Hypothesis Elucidation

This hypothesis posits that differences in antibody consumption rates drive dynamic competition among B cell clones, key to understanding the transition from acute immune response to post-infection sequelae. Antibody response dynamics are primarily determined by antigen load (baseline regulation), clone competition driven by antibody consumption rate (core engine), and immune tolerance (homeostatic constraint). This mechanism is elaborated through a three-phase model:

3.1 Phase One: Antigen Growth Phase – Competition-Driven Clearance and Acute Pathology

In the initial phase of an acute infection, to account for the immune system's rapid response, we propose that the system may preemptively activate a diverse subset of B cells by sensing early infection signals. This action aims to rapidly broaden the initial antibody repertoire, creating a richer "clonal arena" for the subsequent consumption-driven selection. (The validity of this hypothesis does not affect the logical deduction of the model.)

During this phase, the pathogen (antigen) often undergoes exponential growth. This means the effective pathogen load or interaction opportunities ($Ag(t)$) faced by antibodies per unit of time increases dramatically, posing a severe and escalating challenge to the immune system. To understand how antibody consumption drives clonal competition at this stage, we introduce a simplified mathematical concept to describe the consumption probability of a single antibody molecule:

Let P be the basal probability that the antibody effectively binds to a single pathogen unit and is consumed (e.g., cleared as an immune complex) within a unit time.

Assuming each interaction with a pathogen unit is an independent event regarding consumption, the probability that the antibody molecule is not consumed by any of the $Ag(t)$ pathogen units within that unit time (residual probability $P\_residual$) is: $P\_residual = (1 - P)^{Ag(t)}$

Therefore, the total probability that the antibody molecule is consumed within that unit time (total consumption rate $P\_consumed$) is: $P\_consumed = 1 - P\_residual = 1 - (1 - P)^{Ag(t)}$ This formula reveals that even if the basal binding/consumption probability P remains constant, a dramatically increasing antigen load $Ag(t)$ can significantly amplify the total consumption rate $P\_consumed$. We hypothesize that this sharply increased $P\_consumed$ due to rising antigen load is the key signal driving the amplification of the corresponding B cell clone.

We further propose a consumption-driven amplification equation to describe this process:

The amplification rate $dAb/dt$ of a specific antibody clone (or its production capacity) is positively

correlated with the consumption rate P_consumed of its antibody product in the previous cycle:

$$dAb/dt = \gamma \cdot f(P\_consumed) = \gamma \cdot f(1 - (1 - P)^{Ag(t)})$$

Where:

dAb/dt: Represents the instantaneous amplification rate of the specific antibody clone number or its production capacity.

γ: Is a regulatory coefficient reflecting the overall amplification potential of the immune system, resource limitations, influence from competition with other clones, and potentially modulated by the baseline antigen load.

f(...): Is a monotonically increasing function representing how the consumption rate P_consumed translates into amplification signal strength.

P and Ag(t) are defined as above.

The core of this model is that it reveals how the immune system might dynamically match antibody production intensity without relying on direct, precise antigen quantification, but rather by sensing the rate at which antibodies are consumed (driven by the pathogen threat indirectly reflected by Ag(t)).

Competitive Dynamics and Clinical Manifestations:

In this phase, high P_consumed provides a strong amplification signal for antibody clones efficiently binding the pathogen, leading to their preferential and rapid expansion. This quickly increases the proportion of these functional antibodies in the total pool, aiding pathogen control. However, during the early infection stage when resources are relatively abundant and competition is less intense, clonal amplification is primarily driven by individual P_consumed. Therefore, some clones with relatively lower consumption rates, including certain self-reactive clones (if their target self-antigen exposure increases due to inflammation, causing their P_consumed to exceed a certain threshold), might also undergo transient amplification, initiating or exacerbating some self-limiting acute symptoms. Ultimately, as highly efficient specific antibodies are produced in large quantities and dominate through the combined action of consumption-driven and classical affinity maturation pathways, they will utilize resources more effectively, intensify competitive pressure, while immune tolerance mechanisms continue to act, limiting and eventually suppressing the amplification of inefficient or self-reactive clones, leading to the resolution of acute symptoms.

3.2 Phase Two: Antigen Decay Phase – Reversal of Competitive Landscape and Sequelae Risk

As the pathogen load decreases, the consumption rate (P_consumed) of pathogen-specific antibodies plummets, and their consumption-driven amplification advantage vanishes. Concurrently, the overall immune response intensity subsides, and inter-clonal competitive pressure alleviates. Critically, this reversal in the competitive landscape provides a window of opportunity for self-reactive clones that continuously react with self-antigens and maintain a basal consumption rate (P_consumed_self). Their amplification signal (γ · f(P_consumed_self)), though potentially low in absolute terms, gains a relative advantage in the context of reduced competition. Especially if high-affinity self-reactive clones newly generated by SHM exist, or if self-antigen exposure temporarily increases (potentially due to physiological activities like strenuous exercise, environmental stress, or other unknown factors), and if these factors cause the amplification signal strength of the self-reactive clones to breach the immune tolerance threshold, these clones might enter a state of sustained amplification, driving the onset and maintenance of chronic sequelae.

3.3 Phase Three: Homeostasis Re-establishment Phase – Dynamic Remodeling and Reversibility

of Pathological Balance

The pathological equilibrium associated with sequelae is not necessarily irreversible. (Outcomes involving irreversible structural damage and complex immune remodeling mediated by T cell polarization are not the primary focus of this paper.) Encountering a new strong immune stimulus (e.g., re-infection) generates large amounts of high-consumption-rate antibodies (high $P\_consumed\_new$) against the new target. These new clones gain a significant amplification advantage via the consumption-driven mechanism ($dAb/dt\_new = \gamma \cdot f(P\_consumed\_new)$). When competing for limited immune resources (affecting $\gamma$) or proliferation signals, they can effectively suppress ("squeeze out") the self-reactive clones maintaining the sequelae, which have a relatively lower consumption rate ($P\_consumed\_self$). This redistribution of competitive advantage holds the potential to break the original pathological equilibrium, promoting the resolution of sequelae. However, this process also carries risks: when this new strong immune response subsides, it will similarly undergo a competitive landscape shift akin to Phase Two, potentially providing expansion opportunities for new or residual self-reactive clones, thereby inducing, maintaining, or even exacerbating sequelae.

4. Verification Framework

Verifying this hypothesis (consumption-driven dynamic competition of antibody clones) is challenging. The core difficulty lies in precisely and dynamically measuring antibody consumption rates and confirming a direct causal link to B cell clone dynamics within the local tissue microenvironment. Therefore, verification strategies should encompass both direct and indirect approaches.

4.1 Direct Validation Strategy

To directly test this hypothesis, an active-manipulation animal framework can be designed to directly link immunological mechanisms with clinical phenotypes:

First, establish a local autoimmune model with quantifiable pathology dominated by specific B cell clones, and design a controllable "consumption sink" (e.g., a soluble target) for its pathogenic antibody.

To Test the Amplification Effect: Periodically inject the "consumption sink" to create an antibody deficit. The predicted outcome is a rebound increase in the target antibody level, accompanied by a quantifiable exacerbation of pathology.

To Test the Competition Effect: Introduce a strong, unrelated, infection-mimicking immune stimulus locally at the lesion. The prediction is a significant improvement in pathology scores, coupled with suppressed production of the pathogenic antibody, ultimately leading to a decrease in its concentration.

While this framework provides a logically robust validation path, its execution, particularly the precise observation of local dynamics, remains challenging.

4.2 Indirect Validation Strategies

Given the difficulties of direct validation, the following indirect strategies can be explored to gather clues, but interpretation requires caution:

4.2.1 Specific Secondary Infection Studies: Observe sequelae patients experiencing specific "secondary gradient" infection events. The key is that the secondary infection should have a similar route to the primary one and moderate intensity (able to compete effectively without introducing new risks). Analyze the association between this event and changes in symptoms and pathogenic antibodies. Limitations include data acquisition, confounding factor control, and causal inference.

4.2.2 Artificial Intervention Studies (Theoretical Exploration): Design and introduce controllable artificial antigens or immunomodulators to induce a new immune response that competes with pathogenic clones. Key aspects are controllable targeting effectiveness and risk intensity. Analyze the association between intervention effects and changes in symptoms and antibodies. Limitations include design, safety, and ethical challenges.

5. Discussion

5.1 Biological Plausibility and Core Speculation of the "Consumption-Driven Competition Mechanism"

The core challenge of this hypothesis lies in proposing a biologically plausible mechanism explaining how the immune system can sense the "consumption rate" of specific antibodies in vivo and translate this into specific, intensity-adjustable amplification or suppression signals for the B cell clones producing those antibodies. Early serum exchange experiments by Bystryn and others provide crucial clues, strongly suggesting the existence of mechanisms within the body capable of responding to changes in circulating antibody levels and regulating endogenous antibody production, independent of the target antigen. This hints that the immune system possesses internal feedback regulatory capabilities based on product levels, extending beyond classical antigen-driven models.

However, translating this macroscopic phenomenon into specific cellular and molecular pathways, especially achieving the sensing of "consumption rate" rather than just static "concentration" and ensuring clone-specific regulation, remains challenging. Inspired by this and integrating Niels K. Jerne's ideas on self-regulation within the immune system via anti-idiotypic networks[5], we construct a highly speculative core mechanism framework to attempt an explanation for "consumption-driven competition":

Information Presentation: The B cell's "Identity Card": We speculate that B cells continuously internalize and process their own antibody products, presenting the most representative variable region (idiotype) peptides via MHC class II molecules on their surface. This acts as the B cell displaying its "identity"—which specific antibody it produces.

Recognition and Sensing: Specialized T cells as "Sensors" and "Integrators": We envision a class of specialized T helper cells (anti-idiotypic T cells) whose TCRs can specifically recognize the idiotype peptide-MHC II complexes presented by specific B cells. This forms the basis for specific regulatory connections. More critically and speculatively: we assume these T cells also possess some mechanism to sense the actual abundance or trend of change (indirectly reflecting consumption rate) of the corresponding antibody in the environment. This sensing might occur via (non-exclusive) pathways:

Indirect sensing via Antigen Presenting Cells (APCs): APCs (like dendritic cells) continuously capture and present circulating free antibody fragments. If an antibody is abundant in the environment, the density of corresponding peptides on APC surfaces is high, and T cells interacting with them receive a strong "abundance" signal. Conversely, if the antibody is heavily consumed leading to lower concentrations, the signal weakens.

Inefficient direct interaction with free antibody: While TCR recognition is strictly MHC-restricted, we cannot entirely rule out some form of extremely low-affinity but cumulative "interference" or "binding" signal between the free antibody variable region (at very high concentrations) and the TCR (or its complex). The strength of this signal would directly reflect free antibody concentration. MHC restriction might paradoxically prevent T cell saturation by high antibody concentrations, allowing more sensitive perception of relative concentration changes.

Signal Integration and Calibrated Feedback: T cell "Decision" and "Execution": The core function of the anti-idiotypic T cell is to integrate the "identity/request" signal received from the B cell (Signal 1: idiotype peptide-MHC) and the "antibody abundance/demand" signal sensed from the environment (Signal 2). Based on the comparison of these two signals, the T cell makes a decision and provides calibrated feedback to the B cell:

High Demand (Fast Consumption/Low Concentration): When Signal 2 indicates insufficient environmental antibody (e.g., due to rapid consumption lowering concentration), even if Signal 1 intensity is constant, the T cell interprets this as "reinforcements needed," providing strong co-stimulation and cytokines to the B cell, driving its proliferation and differentiation to replenish the antibody.

Sufficient Supply (Slow Consumption/High Concentration): When Signal 2 indicates sufficient or excess environmental antibody, the T cell judges it as "no reinforcements needed" or "suppression required," reducing help to the B cell or potentially delivering inhibitory signals to prevent overproduction.

This mechanism allows antibody production to dynamically match its consumption rate, forming a demand-responsive closed loop.

It must be re-emphasized that the existence of these specialized T cells, their precise sensing mechanisms, and the methods of signal integration and feedback regulation currently lack direct experimental support and represent the most central and weakest speculative link in the mechanism aspect of this hypothesis. Validating or refining this speculative framework is a key direction for future research. Nevertheless, this framework provides a logically possible pathway explaining how the immune system might achieve sensitive responses to the consumption rate of its own products and clone-specific regulation.

Furthermore, to implement such dynamic regulation, the system must address the temporal scaling of its response. We therefore propose a further speculative dual-cycle "central-peripheral" coordinated regulatory model:

Rapid Peripheral Response (Hourly Scale): In the local tissue microenvironment (e.g., a lesion), B cells or plasma cells may react rapidly to the immediate consumption of surrounding antibodies. When local antibodies are depleted through extensive target binding, this could weaken local survival signals for the cells, triggering an "alarm" and transmitting a "supply request" to nearby lymphoid tissues on an hourly or even minute-level timescale.

Central Homeostatic Regulation (Circadian Scale): At the systemic level, the immune system likely operates on a circadian (~24-hour) rhythm. We further speculate that the nightly sleep phase represents the critical window for this central regulation. During sleep, the body's physiological activity is at a minimum, providing a "low-noise" environment. This would allow the immune system to more accurately sense the net consumption rate of circulating antibodies, avoiding interference from physiological fluctuations caused by daytime activities. Extensive research shows that immune cells migrate to lymphoid organs for information processing and memory consolidation during sleep. Therefore, making global calibration decisions during this quiescent period may be crucial for precise amplification or suppression, thereby reducing the risk of erroneously expanding autoreactive clones. This also offers a new mechanistic explanation for the well-known phenomenon that poor sleep quality impairs immune function and can trigger or exacerbate sequelae.

This dual-cycle model depicts an efficient and orderly regulatory system capable of both rapidly responding to local crises and maintaining global production homeostasis, offering a logically plausible dynamic framework for the real-time operation of the "consumption-driven" mechanism.

5.2 Complement and Challenge to Mainstream Immunological Theories

While acknowledging existing theoretical foundations, this hypothesis also offers some complements and challenges:

Extends Clonal Selection Theory: Introduces a feedback regulatory loop based on antibody product consumption, proposing "consumption-driven" as a possible complementary (or even antigen-independent in certain phases) B cell amplification pathway, and explains some immunopathologies from the perspective of "dynamic competitive imbalance" rather than simple "over-activation."

Modifies Static View of Immune Homeostasis: Suggests that the threshold for autoimmune tolerance might not be static but a "dynamic adaptive threshold" influenced by the current overall immune competitive landscape (e.g., the presence of high-consumption antibodies). It also explains how certain autoimmune states might form a "competitive steady-state lock-in" driven by continuous consumption, making spontaneous resolution difficult.

5.3 Validation Challenges and Hypothesis Limitations

As previously stated (Sec. 4), the core challenge of this hypothesis is its testability. Even within the idealized animal model designed in Section 4.1, there remain significant technical difficulties in precisely and dynamically measuring local antibody consumption rates and their immediate impact on B cell production. This makes it complex to establish a direct causal link between systemic antibody concentration changes and local competitive dynamics, which is a major obstacle for current verification efforts.

Major limitations include: the highly speculative nature of the core consumption sensing/transduction mechanism; simplification of complex immune interactions (e.g., other T cell subsets, innate immunity, microenvironment, immunometabolism); the need to test the universality of the model across different pathogens and host backgrounds; and the risk of a single-perspective view on the multifactorial etiology of sequelae. These challenges and limitations collectively point towards directions for future exploration through theoretical refinement and experimental/technological innovation.

5.4 Near-term Feasible Validation Paths and Prospects

Given the technical bottlenecks in direct validation, near-term research should focus on more feasible paths, gradually accumulating indirect evidence and seeking proof-of-concept for key points to support the assessment of this hypothesis.

Accumulating Indirect Correlative Evidence:

Mechanistic Reinterpretation of Existing Therapies: This hypothesis offers a new explanatory framework for certain therapies effective against sequelae but with unclear mechanisms. These treatments may work by introducing a new, highly competitive immune response that reshapes the competitive landscape, thereby suppressing the autoreactive clones that sustain the sequelae. Based on this, hypothesis-driven optimization strategies can be proposed, with comparative clinical outcomes providing evidence.

Systematic Observation of "Infection-Induced Remission of Chronic Diseases": A highly compelling yet often overlooked source of evidence is the observation that some patients experience significant remission, or even resolution, of pre-existing chronic conditions (such as certain autoimmune diseases or allergies) following a strong acute infection. This phenomenon is likely severely underreported due to reporting bias. From our hypothesis's perspective, this is a perfect manifestation of a new, potent immune response successfully "out-competing" the original pathological clones through dynamic competition. Therefore, prospectively and systematically documenting and studying this phenomenon would provide powerful support for the hypothesis from real-world clinical outcomes.

Key Point Proof-of-Concept:

Select sequelae models or specific clinical phenotypes with a relatively good foundation, identifying cases where symptom fluctuations are highly synchronized with inferred indicators of antibody consumption (e.g., target antigen levels, changes in relevant biomarkers).

In these selected models/phenotypes, apply techniques like antibody repertoire lineage tracing and single-cell sequencing to investigate whether there is evidence of competitive abundance shifts (as predicted by the theory) between newly emerged antibody clones triggered by specific events (e.g., natural secondary infection, specific immune intervention) and the putative pathogenic (auto)antibody clones.

While these strategies cannot immediately fully confirm the hypothesis, they hold promise for providing crucial clues to understand the potential role of antibody dynamic competition in post-infection immune dysregulation and guiding subsequent, more in-depth mechanistic studies.